\documentclass[reprint,aps,prb,twocolumn,amsmath,amssymb,showpacs,preprintnumbers]{revtex4-2}

\usepackage{graphicx}[]
\usepackage{dcolumn}
\usepackage{bm}
\usepackage{multirow}

\usepackage{braket}
\usepackage{color}
\usepackage{ulem}
\usepackage{url}
\usepackage{hyperref}

\begin{document}

\title{
Three-sublattice antiferro-type and ferri-type skyrmion crystals in centrosymmetric magnets
}
\author{Satoru Hayami}
\affiliation{
Graduate School of Science, Hokkaido University, Sapporo 060-0810, Japan
}

\begin{abstract}
We numerically investigate the stability of the skyrmion crystals in a centrosymmetric lattice structure by focusing on the role of magnetic frustration arising from the multi-sublattice degree of freedom. 
By analyzing an effective three-sublattice spin model with the antiferromagnetic exchange interaction between different sublattices and the momentum-resolved exchange interaction between the same sublattice based on the simulated annealing, we find that various skyrmion crystal phases with sublattice-dependent skyrmion numbers are stabilized with and without an external magnetic field depending on the model parameters: two ferro-type skyrmion crystal phases with the skyrmion numbers of one and two, two ferri-type skyrmion crystal phases, and one antiferro-type skyrmion crystal phase. 
Our results indicate that the multi-sublattice degree of freedom brings a further intriguing possibility of inducing exotic skyrmion crystal phases. 
\end{abstract}

\maketitle

\section{Introduction}

A magnetic skyrmion crystal (SkX) with nontrivial topological properties has been extensively studied since its direct experimental observation in 2009~\cite{Muhlbauer_2009skyrmion,yu2010real}. 
Although it was originally discovered in noncentrosymmetric magnets~\cite{Muhlbauer_2009skyrmion, yu2010real, seki2012observation, Adams2012}, where the competition between the ferromagnetic exchange interaction and the Dzyaloshinskii-Moriya (DM) interaction~\cite{dzyaloshinsky1958thermodynamic,moriya1960anisotropic} plays an important role~\cite{rossler2006spontaneous, Binz_PhysRevLett.96.207202,  Binz_PhysRevB.74.214408, Yi_PhysRevB.80.054416}, its appearance in centrosymmetric magnets has been clarified through the observations for various bulk materials, such as Gd$_2$PdSi$_3$~\cite{kurumaji2019skyrmion, Paddison_PhysRevLett.129.137202, Bouaziz_PhysRevLett.128.157206, nomoto2023ab}, Gd$_3$Ru$_4$Al$_{12}$~\cite{hirschberger2019skyrmion,Hirschberger_10.1088/1367-2630/abdef9}, GdRu$_2$Si$_2$~\cite{khanh2020nanometric,Yasui2020imaging, khanh2022zoology, Matsuyama_PhysRevB.107.104421, hayami2023widely, Wood_PhysRevB.107.L180402} and EuAl$_4$~\cite{takagi2022square,Zhu_PhysRevB.105.014423, hayami2023orthorhombic, Gen_PhysRevB.107.L020410}. 
Simultaneously, a variety of SkXs without relying on the DM interaction have been theoretically studied for various lattice structures~\cite{hayami2021topological}, such as square~\cite{Hayami_PhysRevB.103.024439, Utesov_PhysRevB.103.064414, Wang_PhysRevB.103.104408, Hayami_PhysRevB.105.174437}, triangular~\cite{Okubo_PhysRevLett.108.017206, leonov2015multiply, Hayami_PhysRevB.103.054422, Utesov_PhysRevB.105.054435, Eto_PhysRevLett.129.017201, wang2021skyrmion}, and cubic lattices~\cite{hayami2021field}. 
Since the SkX in centrosymmetric magnets tends to have a short magnetic modulation period, it has a great advantage to energy-efficient devices based on high-density topological objects, which is promising for future spintronics applications. 

Meanwhile, exploring further intriguing topological magnetic states is one of the challenging issues in condensed matter physics~\cite{gobel2021beyond}. 
In this context, a plethora of spin textures have been proposed and investigated, such as a meron-antimeron crystal~\cite{Lin_PhysRevB.91.224407, yu2018transformation, Hayami_PhysRevB.104.094425}, skyrmionium~\cite{Zhang_PhysRevB.94.094420, zhang2018real}, hedgehog~\cite{Kanazawa_PhysRevB.86.134425, kanazawa2016critical, Ishiwata_PhysRevB.101.134406, Okumura_PhysRevB.101.144416}, and hopfion~\cite{Liu_PhysRevB.98.174437, Liu_PhysRevLett.124.127204, kent2021creation}. 
Among them, we focus on the role of the multi-sublattice degree of freedom under lattice structures in the stabilization of topological magnetic states. 
The most typical example is an antiferro-type SkX (AF-SkX) in two-sublattice systems, where the skyrmion number corresponding to the topological charge exhibits the opposite sign between two sublattices so that it is canceled out in the whole system~\cite{Bogdanov_PhysRevB.66.214410, buhl2017topological, Gobel_PhysRevB.96.060406, Akosa_PhysRevLett.121.097204}. 
The realization of the AF SkX has been clarified in the presence of an external staggered magnetic field~\cite{Gobel_PhysRevB.96.060406}, uniform out-of-plane magnetic field~\cite{Yambe_PhysRevB.107.014417}, and uniform in-plane magnetic field~\cite{Hayami_doi:10.7566/JPSJ.92.084702}. 
Especially, not only the AF SkX but also the ferri-type SkX (Ferri-SkX) with the different skyrmion numbers for different sublattices have been found in itinerant honeycomb magnets with the two-sublattice structure~\cite{Yambe_PhysRevB.107.014417}. 
Based on these observations, a natural question arises: is it possible to realize the AF SkX in odd-number sublattice systems, e.g., the three-sublattice system? 
Although the three-sublattice SkX have been investigated in various theoretical models~\cite{Rosales_PhysRevB.92.214439, Diaz_hysRevLett.122.187203, Osorio_PhysRevB.96.024404, liu2020theoretical, Tome_PhysRevB.103.L020403}, no SkX with different skyrmion numbers at each sublattice but without the skyrmion number in the total system was reported except for a trilayer model with a layer-dependent DM interaction~\cite{Hayami_PhysRevB.105.184426}. 

In the present study, we investigate the stabilization of the three-sublattice AF SkX in centrosymmetric magnets. 
By performing the simulated annealing for an effective spin model with the momentum-resolved interaction between the same sublattice and antiferromagnetic interaction between the different sublattices, we construct a ground-state phase diagram in the fundamental three-sublattice magnetic system. 
We find that the AF-SkX phase appears for the intermediate magnetic field, where two out of three sublattices exhibit the spin textures with the skyrmion number of $+1$ ($-1$), while the remaining sublattice shows that with the skyrmion number of $-2$ ($+2$) so as to vanish the skyrmion number in the whole system. 
Furthermore, we find rich SkX phases with different skyrmion numbers depending on the inter-sublattice antiferromagnetic interaction and the magnetic field: two types of the ferro-type SkX (F-SkX) and two types of the Ferri-SkX. 
Our results indicate that materials with multi-sublattice structures can host exotic topological spin textures that are distinct from those in the single-sublattice ones. 

The rest of the paper is organized as follows. 
In Sec.~\ref{sec: Model}, we introduce an effective spin model and the method based on the simulated annealing. 
Then, we show the magnetic phase diagram in Sec.~\ref{sec: Results}. 
We discuss the detailed spin and chirality textures in the obtained phases including the AF-SkX. 
A summary of the results is presented in Sec.~\ref{sec: Summary}. 

\section{Model and method}
\label{sec: Model}

\begin{figure}[tb!]
\begin{center}
\includegraphics[width=1.0\hsize]{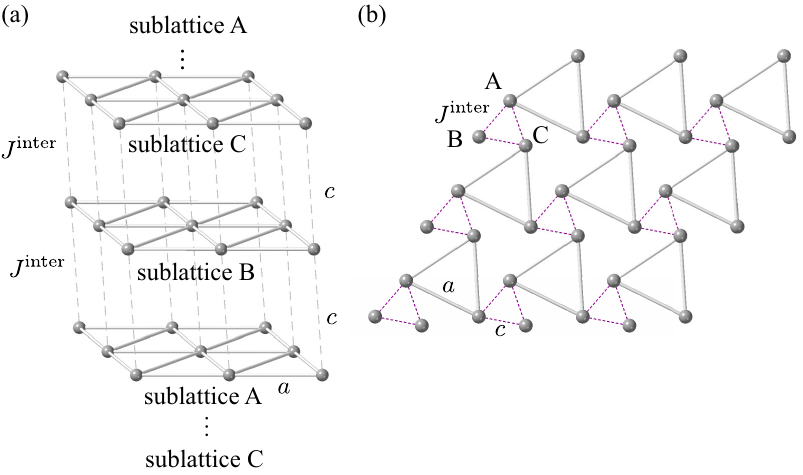} 
\caption{
\label{fig: lattice} 
Examples of three-sublattice structures with sublattices A--C: (a) trilayer and (b) trigonal structures. 
$a$ and $c$ are the length of each bond. 
$J^{\rm inter}$ stands for the exchange interaction between the different sublattices. 
}
\end{center}
\end{figure}

In order to investigate the instability toward the AF SkX in the three-sublattice system, we consider an effective spin model on a lattice consisting of three sublattices A--C under a trigonal lattice structure, which is given by 
\begin{align}
\label{eq: Ham}
\mathcal{H}=\sum_{\eta}\mathcal{H}^{\rm intra}_{\eta}+\mathcal{H}^{\rm inter}+\mathcal{H}^{{\rm Z}}, 
\end{align}
where 
\begin{align}
\label{eq: Ham_intra}
\mathcal{H}^{\rm intra}_{\eta}=&  2\sum_{\nu}  \left[- J\bm{S}_{\eta \bm{Q}_\nu} \cdot \bm{S}_{\eta-\bm{Q}_\nu}+ \frac{K}{N}(\bm{S}_{\eta \bm{Q}_\nu} \cdot \bm{S}_{\eta-\bm{Q}_\nu})^2  \right], \\
\label{eq: Ham_inter}
\mathcal{H}^{\rm inter}=& J^{\rm inter} \sum_{\langle i, j \rangle } \bm{S}_i \cdot \bm{S}_{j},\\
\label{eq: Ham_Zeeman}
\mathcal{H}^{{\rm Z}}=&-H \sum_i S_i^z. 
\end{align}
The first term in Eq.~(\ref{eq: Ham}), $\mathcal{H}^{\rm intra}_{\eta}$, represents the Hamiltonian for sublattice $\eta=$ A--C, which includes the bilinear and biquadratic exchange interactions with the coupling constants $J$ and $K$, respectively, in momentum space; $\bm{S}_{\eta \bm{Q}_\nu}$ is the spin moment at the $\bm{Q}_\nu$ component for sublattice $\eta$; $\bm{S}_{\eta \bm{Q}_\nu}$ is related to the real-space spin $\bm{S}_i$ at site $i$ by the Fourier transformation; we fix the spin length as $|\bm{S}_i| = 1$. 
We consider the dominant interactions in momentum space by choosing several $\bm{Q}_\nu$ in the Brillouin zone; we take the contributions from the symmetry-equivalent wave vectors by assuming the threefold rotational symmetry in the system: $\bm{Q}_1=(\pi/3,0, 0)$, $\bm{Q}_2=(-\pi/6, \sqrt{3}\pi/6, 0)$, and $\bm{Q}_3=(-\pi/6, -\sqrt{3}\pi/6, 0)$ (we set the lattice constant as unity as will be discussed below), and the prefactor 2 in Eq.~(\ref{eq: Ham_intra}) represents the contribution from $-\bm{Q}_1$, $-\bm{Q}_2$, and $-\bm{Q}_3$. 
The simplification to extract the dominant interactions in momentum space is justified when the ground-state spin configurations are calculated, since the contributions to the ground-state internal energy are well dominated by specific wave vectors.
Microscopically, this Hamiltonian in Eq.~(\ref{eq: Ham_intra}) is derived as an effective spin model of the Kondo lattice model in the weak-coupling regime, where the bare susceptibility of itinerant electrons shows maxima at $\bm{Q}_1$--$\bm{Q}_3$~\cite{Hayami_PhysRevB.95.224424, Yambe_PhysRevB.106.174437}; the first term corresponds to the lowest contribution in the perturbation expansion, i.e., Ruderman-Kittel-Kasuya-Yosida interaction~\cite{Ruderman, Kasuya, Yosida1957}, and the second term corresponds to the second-lowest contribution, the latter of which tends to induce various multiple-$Q$ instabilities when its sign is positive~\cite{Akagi_PhysRevLett.108.096401, Hayami_PhysRevB.90.060402, Hayami_PhysRevB.95.224424}, such as the double-$Q$ SkX found in GdRu$_2$Si$_2$~\cite{khanh2022zoology, hayami2023widely} and triple-$Q$ vortex crystal in Y$_3$Co$_8$Sn$_4$~\cite{takagi2018multiple}. 
We set $J$ as the energy unit of the model Hamiltonian $\mathcal{H}$ and treat $K>0$ as the phenomenological parameter. 

The second term in Eq.~(\ref{eq: Ham}), $\mathcal{H}^{\rm inter}$, denotes the inter-sublattice Hamiltonian. 
We consider the antiferromagnetic exchange interaction ($J^{\rm inter}>0$) for all pairs of different sublattices by implicitly considering a trilayer triangular system in Fig.~\ref{fig: lattice}(a); we set $a=c=1$ for simplicity. 
Meanwhile, it is noted that the model with the interactions in Eqs.~(\ref{eq: Ham_intra}) and (\ref{eq: Ham_inter}) is mapped onto that in a different three-sublattice system; a two-dimensional trigonal lattice in Fig.~\ref{fig: lattice}(b) is geometrically equivalent to the trilayer lattice in Fig.~\ref{fig: lattice}(a) within the nearest-neighbor inter-sublattice interaction. 
They are distinguished by considering further-neighbor interactions in $\mathcal{H}^{\rm inter}$ and magnetic anisotropic interactions in $\mathcal{H}^{\rm intra}_{\eta}$, which are not taken into account in the present model in order to focus on the role of the competition between $K$ and $J^{\rm inter}$~\cite{Hayami_PhysRevB.105.014408, Hayami_PhysRevB.105.184426, hayami2022square}. 
Thus, the model in Eq.~(\ref{eq: Ham}) is the most fundamental model to describe the multiple-$Q$ instabilities in a three-sublattice system. 
The last term in Eq.~(\ref{eq: Ham}), $\mathcal{H}^{{\rm Z}}$, represents the Zeeman coupling in the presence of an external magnetic field along the $z$ direction.

We calculate the magnetic phase diagram at low temperatures by means of the simulated annealing for the model in Eq.~(\ref{eq: Ham}) on the trilayer triangular lattice with the total number of spins $N=3 \times L^2$ for $L=12$ under the periodic boundary conditions in all the three direction.  
The details of the simulated annealing are as follows: 
First, we choose a random spin configuration at high temperatures as the initial temperature $T_0=0.1$-$1.0$. 
Then, we gradually reduce the temperature with a cooling rate $T_{n+1}=\alpha T_n$ until the temperature reaches the final temperature at $T=0.001$, where $T_n$ is the temperature in the $n$th step and $\alpha=0.99999-0.999999$. 
While decreasing the temperature, we perform standard Metropolis local updates in real space at each temperature. 
We totally perform $10^5$-$10^6$ Monte Carlo sweeps for annealing and measurements at the final temperature.
To evade a metastable solution near the phase boundaries, we also perform the simulations by starting from the spin configurations obtained at low temperatures. 

The magnetic phases obtained by the simulated annealing are classified into their spin and chirality textures. 
In the spin textures, we calculate the spin structure factor per layer $\eta$ as 
\begin{align}
\label{eq:Sq}
S_{\eta}^\alpha(\bm{q})= \frac{1}{L^2} \sum_{i,j \in \eta} S^{\alpha}_i S^{\alpha}_j e^{i\bm{q}\cdot (\bm{r}_i-\bm{r}_j)}, 
\end{align}
where $\alpha=x,y,z$. 
We also calculate magnetic moments at the $\bm{Q}_\nu$ component as 
\begin{align}
m^\alpha_{\eta \bm{Q}_\nu}=\sqrt{\frac{S_{\eta}^\alpha(\bm{Q}_\nu)}{N}}.  
\end{align}
The total spin component of $m^\alpha_{\eta \bm{Q}_\nu}$ is defined as $m_{\eta \bm{Q}_\nu}=\sqrt{(m^x_{\eta \bm{Q}_\nu})^2+(m^y_{\eta \bm{Q}_\nu})^2+(m^z_{\eta \bm{Q}_\nu})^2}$. 
The net magnetization per layer is given by
\begin{align}
M^\alpha_{\eta}=\frac{1}{L^2}\sum_{i \in \eta}S^{\alpha}_{i}.
\end{align}

In the chirality texture, the spin scalar chirality is represented by 
\begin{align}
\label{eq: chirality}
\chi^{\rm sc}_{\eta} &= \frac{1}{L^2} \sum_{\bm{R}\in \eta}\chi_{\bm{R}},\\
\chi_{\bm{R}}&= \bm{S}_{i} \cdot (\bm{S}_j \times \bm{S}_k),
\end{align}
where $\chi_{\bm{R}}$ represents the local scalar chirality at the position vector $\bm{R}$, which is located at the centers of upward and downward triangles with the vertices $i$, $j$, and $k$ within the same sublattice in the counterclockwise order. 
By using $\chi_{\bm{R}}$, the skyrmion number per layer is calculated as 
\begin{align}
\label{eq:nsk}
n^\eta_\mathrm{sk} = \frac{1}{4\pi N_\mathrm{m}}\left\langle \sum_{\bm{R}\in \eta} \Omega^{\eta}_{\bm{R}} \right\rangle,
\end{align}
where $N_{\rm m}$ is the number of the magnetic unit cell and $\Omega^{\eta}_{\bm{R}}$ is the skyrmion density for layer $\eta$~\cite{BERG1981412}: 
\begin{align}
\tan\left(\frac{\Omega^{\eta}_{\bm{R}}}{2}\right) = \frac{\chi_{\bm{R}}}{1+\bm{S}_i\cdot\bm{S}_j+\bm{S}_j\cdot\bm{S}_k+\bm{S}_k\cdot\bm{S}_i} .
\end{align}
For example, the SkX with the vortex-like winding of spins around the skyrmion core leads to $n^{\eta}_\mathrm{sk}=-1$, while the anti-SkX with the anti-vortex-like winding of spins leads to $n^{\eta}_\mathrm{sk}=+1$.

\section{Results}
\label{sec: Results}

\begin{figure}[t!]
\begin{center}
\includegraphics[width=0.6\hsize]{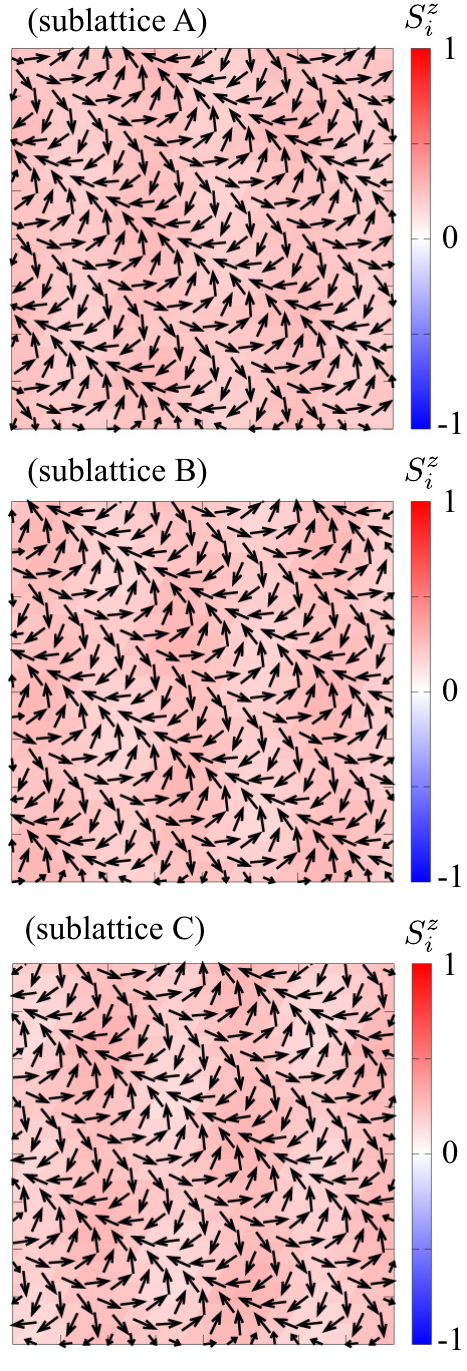} 
\caption{
\label{fig: spin_K=0} 
Real-space spin configurations of the single-$Q$ conical state at $K=0$, $J^{\rm inter}=0.85$ and $H=1$ for sublattice A (upper panel), sublattice B (middle panel), and sublattice C (lower panel).  
The arrows represent the $xy$ components of the spin moment and the color shows the $z$ component. 
}
\end{center}
\end{figure}

There are three independent parameters in the model in Eq.~(\ref{eq: Ham}): $K$, $J^{\rm inter}$, and $H$. 
First, let us discuss the result at $K=0$. 
In this case, the ground state becomes the single-$Q$ state irrespective of $J^{\rm inter}$ and $H$. 
We show the spin configurations for sublattices A--C at $J^{\rm inter}=0.85$ and $H=1$ as an example in Fig.~\ref{fig: spin_K=0}. 
Here and hereafter, we copied the $12\times 12$ spin configurations in each layer, which was obtained by the simulated annealing, so as to satisfy the periodic boundary condition for better visibility. 
The spin configurations in all the layers show the single-$Q$ conical spiral structure with the $\bm{Q}_3$ component, although the phase among the spiral waves is different from each other; the spins at the same $xy$ position from the 120$^{\circ}$ antiferromagnetic structure so that the energy by the interlayer antiferromagnetic exchange $J^{\rm inter}$ is minimized. 
In the end, there is no multiple-$Q$ instability in the absence of $K$.

\begin{figure}[tb!]
\begin{center}
\includegraphics[width=1.0\hsize]{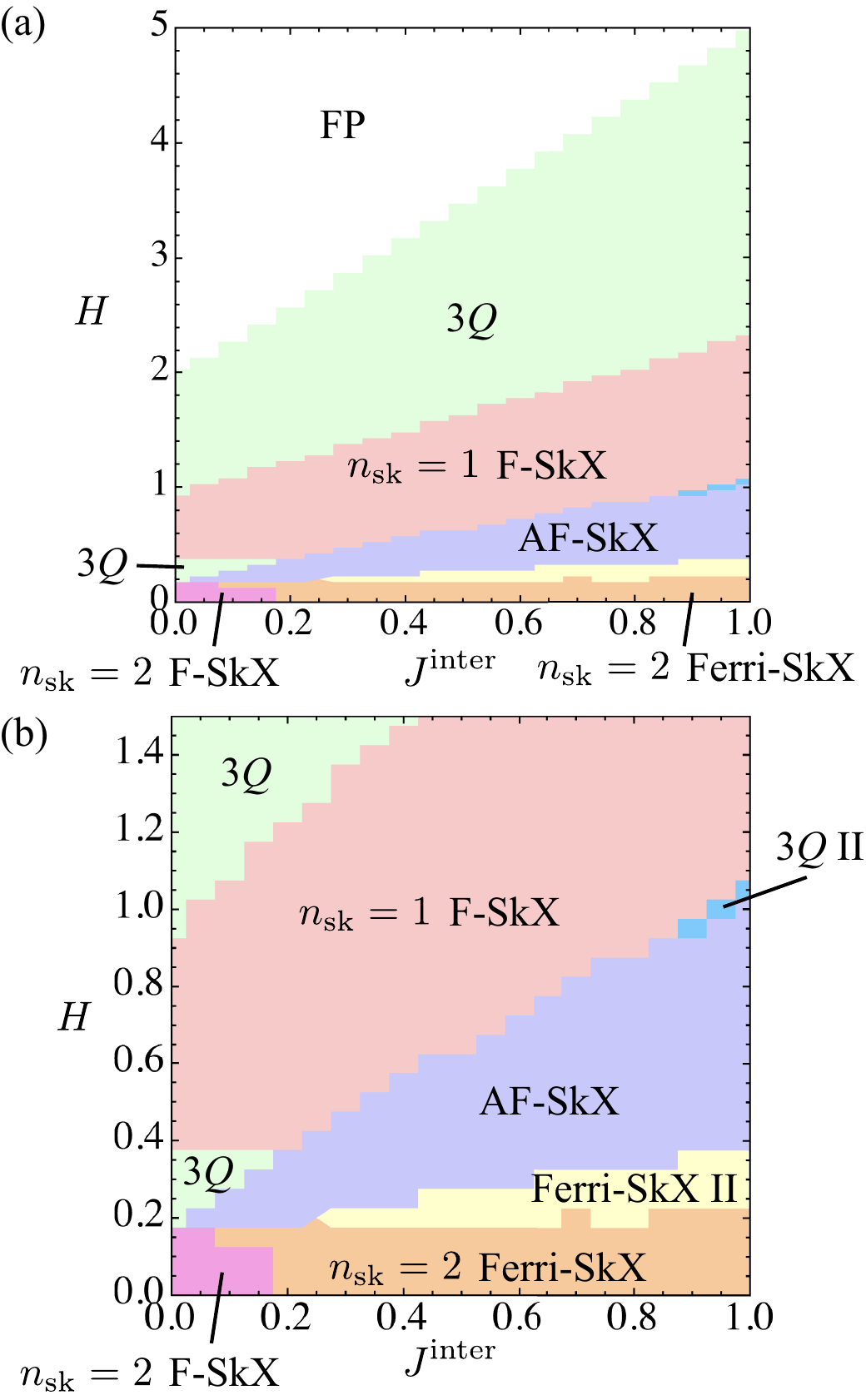} 
\caption{
\label{fig: PD} 
(a) Magnetic phase diagram with varying the inter-sublattice interaction $J^{\rm inter}$ and the external magnetic field $H$ at $K=0.3$, which is obtained by performing the simulated annealing. 
F-SkX, AF-SkX, Ferri-SkX, 3$Q$, and FP represent the ferro-type skyrmion crystal, antiferro-type skyrmion crystal, ferri-type skyrmion crystal, triple-$Q$ state, and fully-polarized state, respectively. 
(b) Enlarged figure of (a). 
}
\end{center}
\end{figure}

\begin{figure}[tb!]
\begin{center}
\includegraphics[width=1.0\hsize]{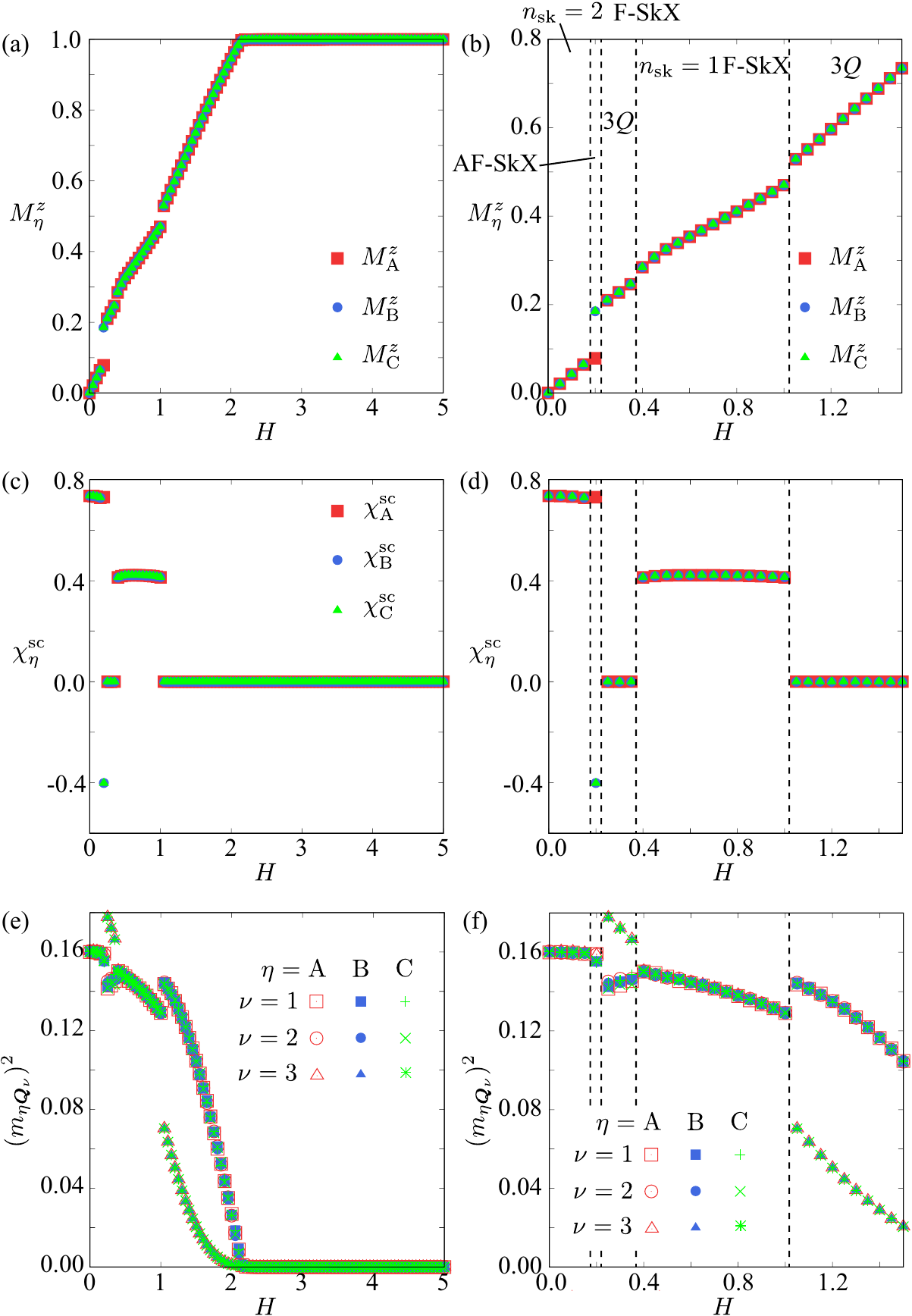} 
\caption{
\label{fig: mq_0.05} 
$H$ dependence of (a,b) the magnetization $M^{z}_\eta$, (c,d) the scalar chirality $\chi^{\rm sc}_{\eta}$, and (e,f) the squared momentum-resolved magnetic moment $(m_{\eta \bm{Q}_\nu})^2$ for sublattices $\eta=$ A, B, and C at $J^{\rm inter}=0.05$. 
(b), (d), and (f) represent the enlarged figures of (a), (c), and (e) in the low-field region, respectively. 
The vertical dashed lines in (b,d,f) represent the phase boundaries between different phases, where the corresponding phases are presented above in (b). 
}
\end{center}
\end{figure}

\begin{figure}[tb!]
\begin{center}
\includegraphics[width=1.0\hsize]{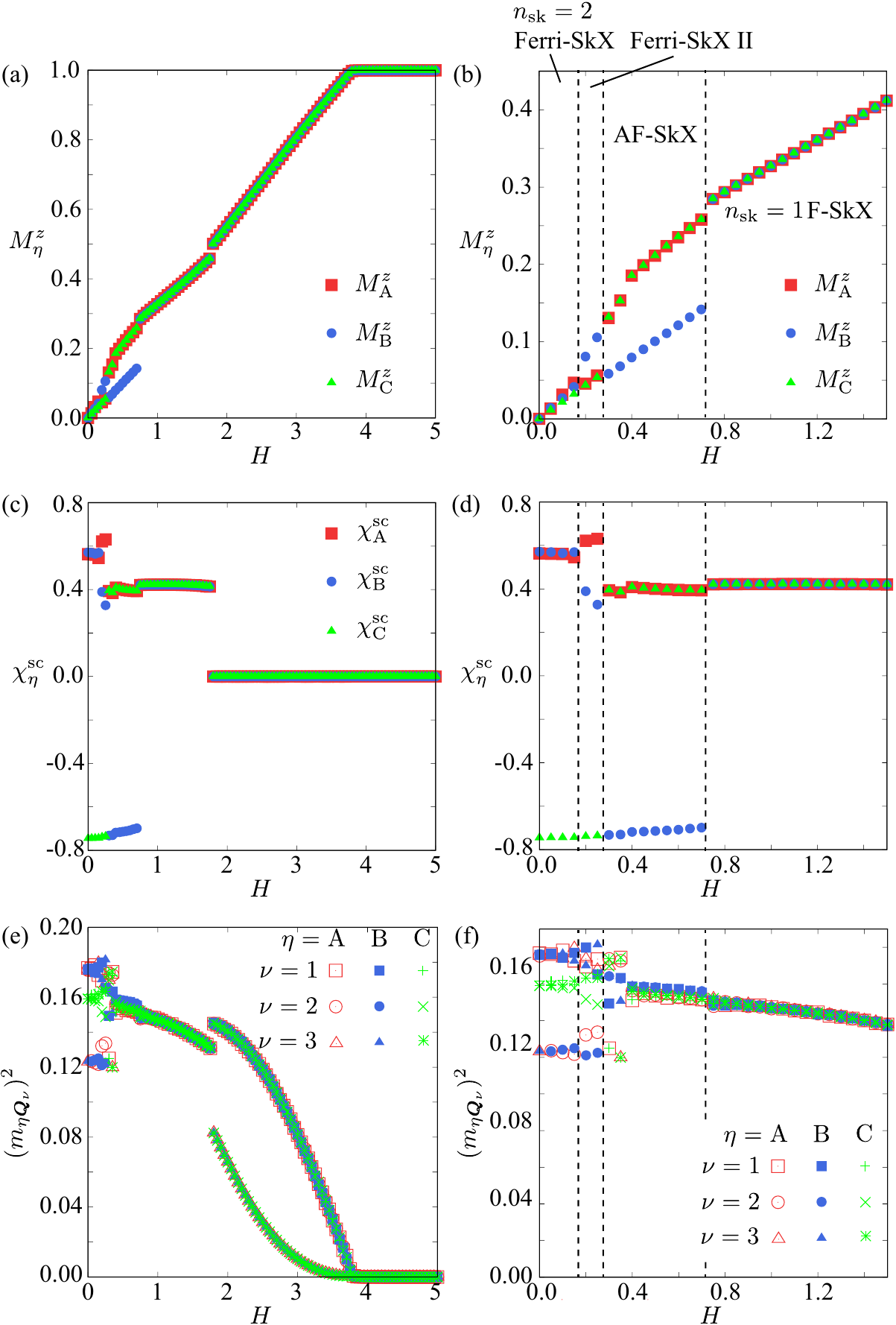} 
\caption{
\label{fig: mq_0.6} 
The same plot as Fig.~\ref{fig: mq_0.05} with $J^{\rm inter}=0.6$. 
}
\end{center}
\end{figure}

\begin{figure}[tb!]
\begin{center}
\includegraphics[width=1.0\hsize]{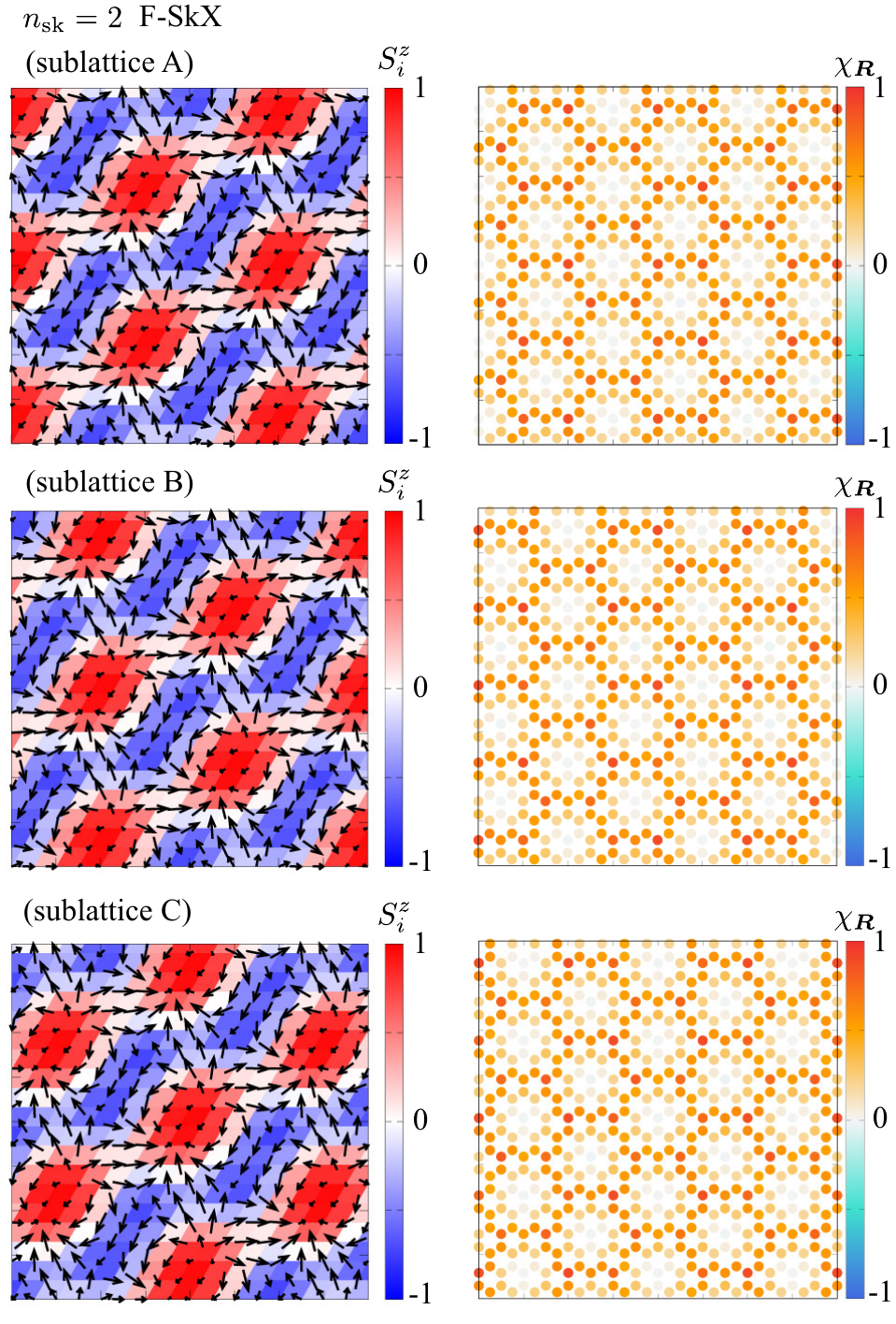} 
\caption{
\label{fig: spin1} 
Real-space spin (left panel) and scalar chirality (right panel) configurations of the $n_{\rm sk}=2$ F-SkX with $(n_{\rm sk}^{\rm A}, n_{\rm sk}^{\rm B}, n_{\rm sk}^{\rm C})=(2,2,2)$ at $J^{\rm inter}=0.05$ and $H=0.1$ for sublattice A (upper panel), sublattice B (middle panel), and sublattice C (lower panel).  
In the left panels, the arrows represent the $xy$ components of the spin moment and the color shows the $z$ component. 
}
\end{center}
\end{figure}

\begin{figure}[tb!]
\begin{center}
\includegraphics[width=1.0\hsize]{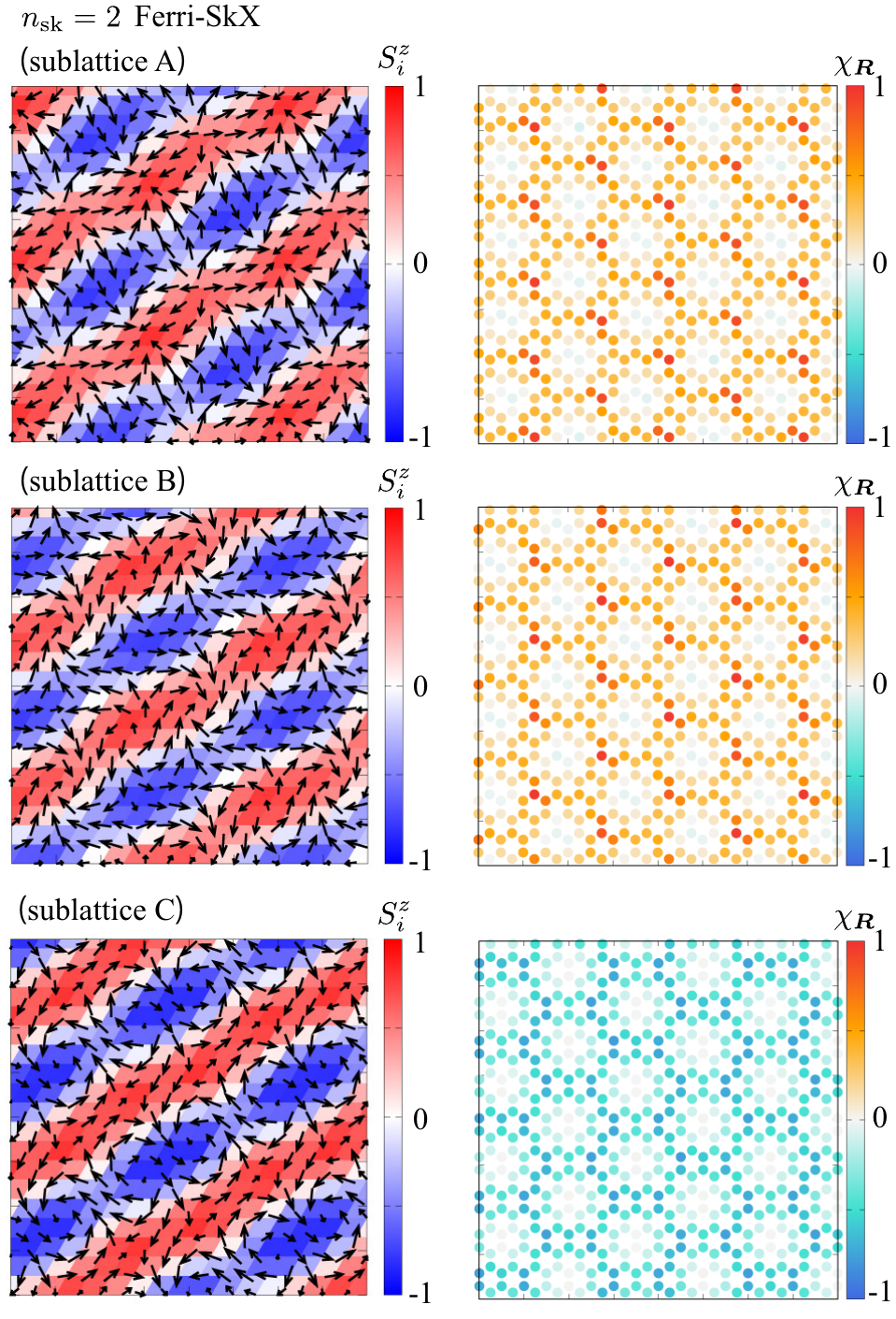} 
\caption{
\label{fig: spin2} 
Real-space spin (left panel) and scalar chirality (right panel) configurations of the $n_{\rm sk}=2$ Ferri-SkX state with $(n_{\rm sk}^{\rm A}, n_{\rm sk}^{\rm B}, n_{\rm sk}^{\rm C})=(2,2,-2)$ at $J^{\rm inter}=0.6$ and $H=0.1$ for sublattice A (upper panel), sublattice B (middle panel), and sublattice C (lower panel).  
In the left panels, the arrows represent the $xy$ components of the spin moment and the color shows the $z$ component. 
}
\end{center}
\end{figure}

\begin{figure}[tb!]
\begin{center}
\includegraphics[width=1.0\hsize]{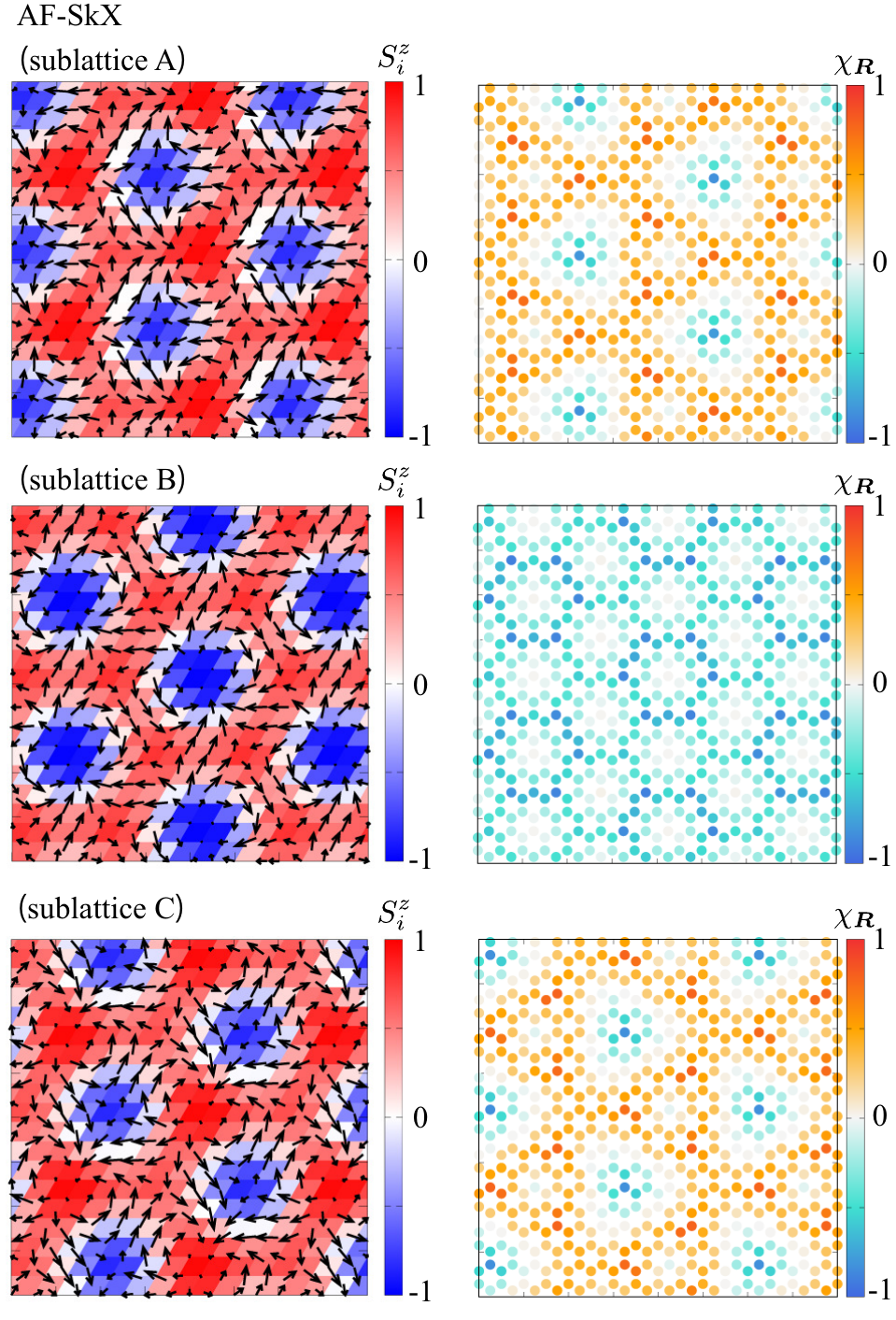} 
\caption{
\label{fig: spin3} 
Real-space spin (left panel) and scalar chirality (right panel) configurations of the AF-SkX state with $(n_{\rm sk}^{\rm A}, n_{\rm sk}^{\rm B}, n_{\rm sk}^{\rm C})=(1,-2,1)$ at $J^{\rm inter}=0.6$ and $H=0.6$ for sublattice A (upper panel), sublattice B (middle panel), and sublattice C (lower panel).  
In the left panels, the arrows represent the $xy$ components of the spin moment and the color shows the $z$ component. 
}
\end{center}
\end{figure}

\begin{figure}[tb!]
\begin{center}
\includegraphics[width=1.0\hsize]{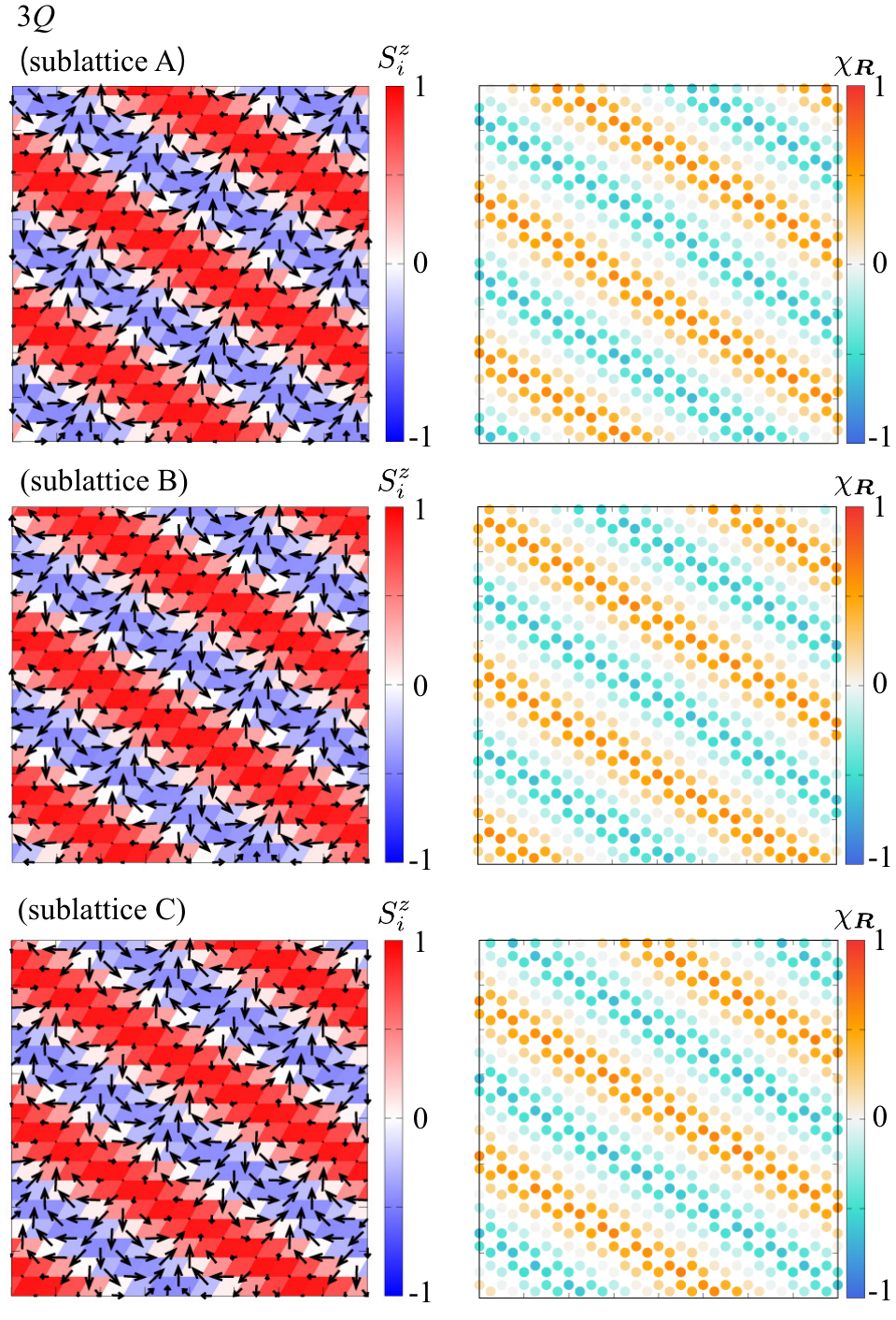} 
\caption{
\label{fig: spin4} 
Real-space spin (left panel) and scalar chirality (right panel) configurations of the 3$Q$ state with $(n_{\rm sk}^{\rm A}, n_{\rm sk}^{\rm B}, n_{\rm sk}^{\rm C})=(0,0,0)$ at $J^{\rm inter}=0.05$ and $H=0.3$ for sublattice A (upper panel), sublattice B (middle panel), and sublattice C (lower panel).  
In the left panels, the arrows represent the $xy$ components of the spin moment and the color shows the $z$ component. 
}
\end{center}
\end{figure}

\begin{figure}[tb!]
\begin{center}
\includegraphics[width=1.0\hsize]{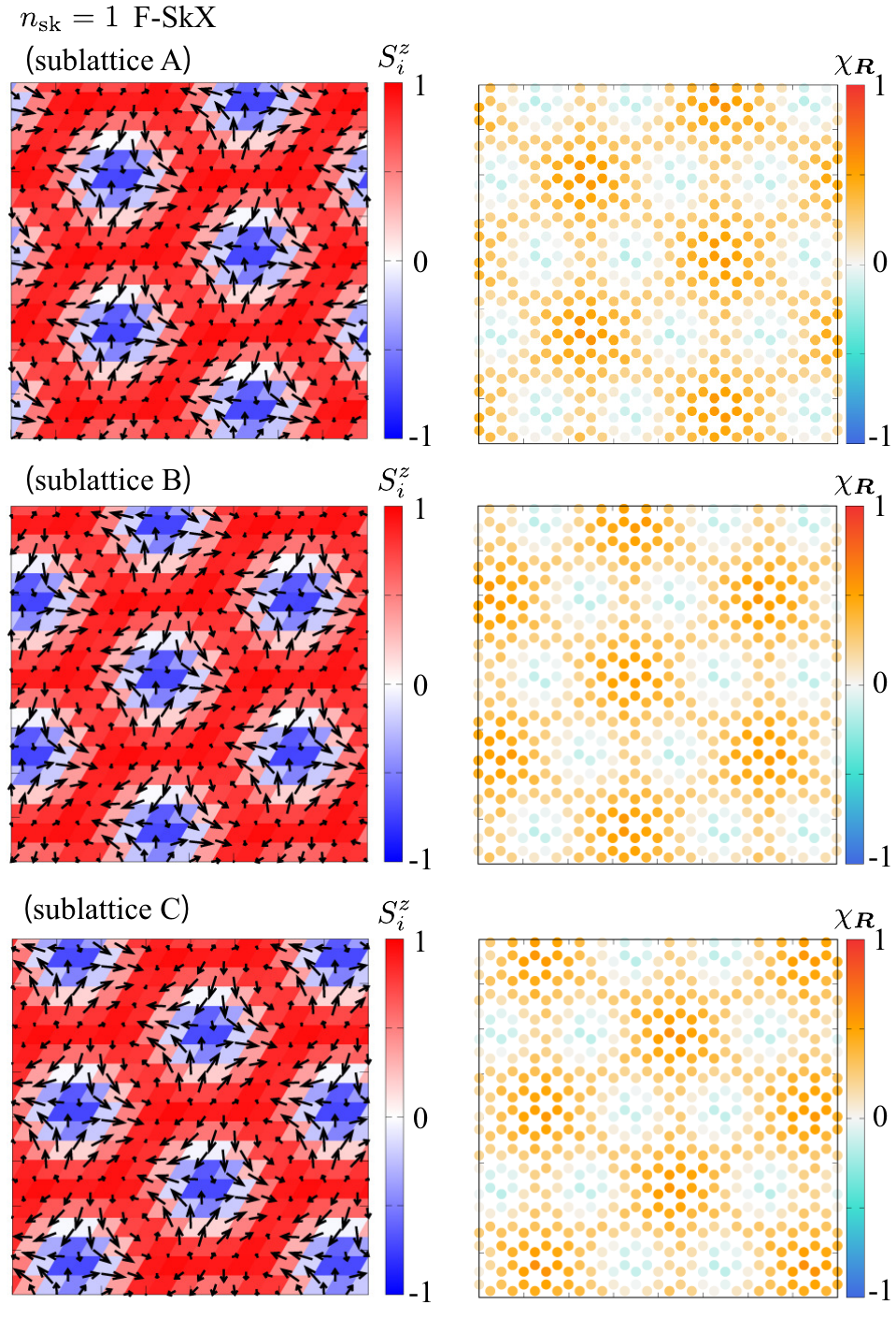} 
\caption{
\label{fig: spin5} 
Real-space spin (left panel) and scalar chirality (right panel) configurations of the $n_{\rm sk}=1$ F-SkX state with $(n_{\rm sk}^{\rm A}, n_{\rm sk}^{\rm B}, n_{\rm sk}^{\rm C})=(1,1,1)$ at $J^{\rm inter}=0.05$ and $H=0.8$ for sublattice A (upper panel), sublattice B (middle panel), and sublattice C (lower panel).  
In the left panels, the arrows represent the $xy$ components of the spin moment and the color shows the $z$ component. 
}
\end{center}
\end{figure}

\begin{figure}[tb!]
\begin{center}
\includegraphics[width=1.0\hsize]{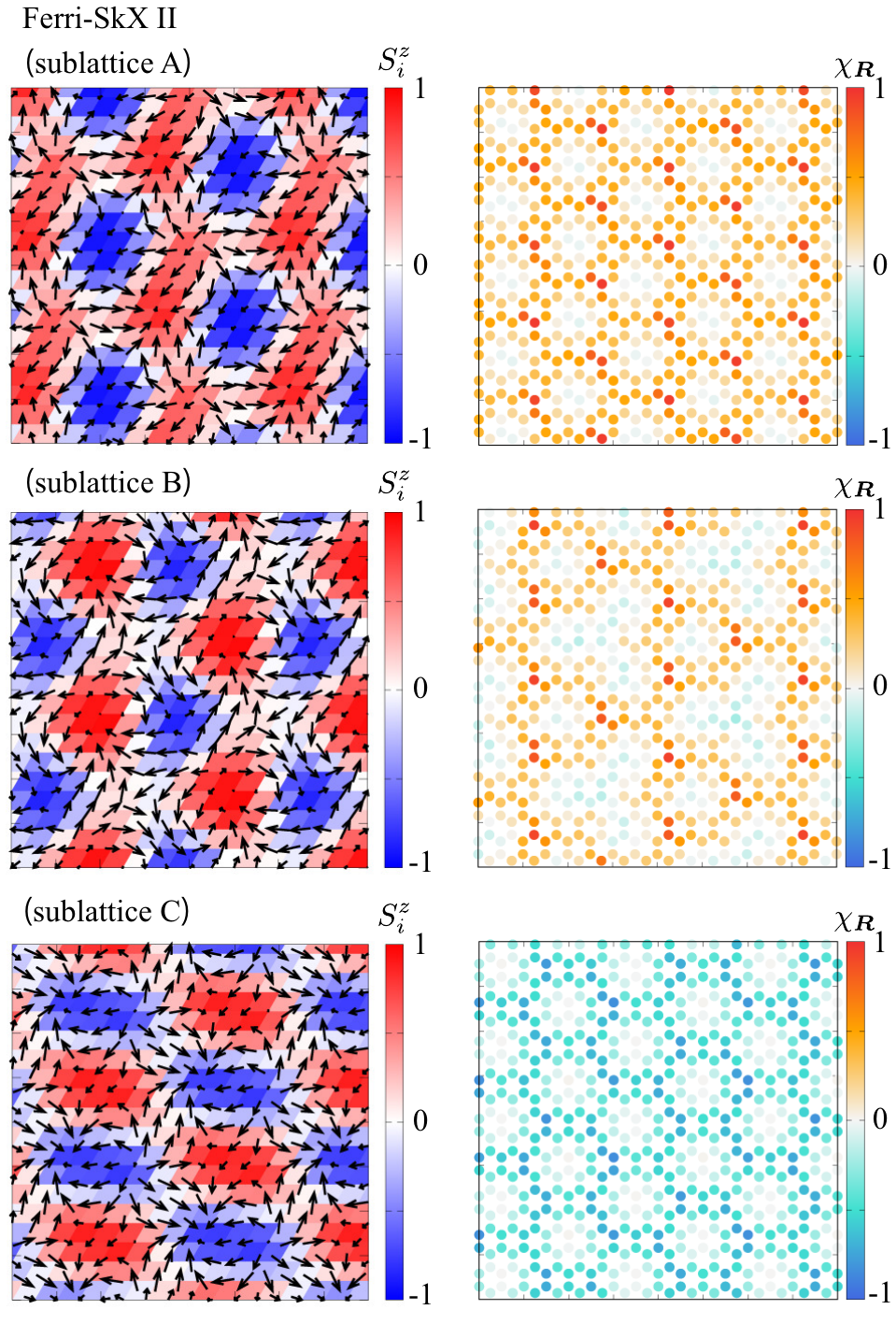} 
\caption{
\label{fig: spin6} 
Real-space spin (left panel) and scalar chirality (right panel) configurations of the Ferri-SkX II with $(n_{\rm sk}^{\rm A}, n_{\rm sk}^{\rm B}, n_{\rm sk}^{\rm C})=(2,1,-2)$ at $J^{\rm inter}=0.6$ and $H=0.2$ for sublattice A (upper panel), sublattice B (middle panel), and sublattice C (lower panel).  
In the left panels, the arrows represent the $xy$ components of the spin moment and the color shows the $z$ component. 
}
\end{center}
\end{figure}

Next, let us consider the effect of $K$. 
Figure~\ref{fig: PD}(a) shows the magnetic phase diagram obtained by the simulated annealing for nonzero $K=0.3$. 
Depending on $J^{\rm inter}$ and $H$, a variety of multiple-$Q$ states are stabilized in the phase diagram. 
There are totally eight phases, among which five types of the SkX phases are obtained; see also Fig.~\ref{fig: PD}(b), which is an enlarged figure of Fig.~\ref{fig: PD}(a).  
We discuss the behaviors of the spin and chirality quantities for obtained phases by referring their $H$ dependence in Figs.~\ref{fig: mq_0.05} and \ref{fig: mq_0.6} and real-space configurations in Figs.~\ref{fig: spin1}, \ref{fig: spin2}, \ref{fig: spin3}, \ref{fig: spin4}, \ref{fig: spin5}, and \ref{fig: spin6}. 
In Figs.~\ref{fig: mq_0.05} and \ref{fig: mq_0.6}, we show the data by appropriately sorting $(m^\alpha_{\eta \bm{Q}_\nu})^2$ for better readability. 
Meanwhile, we show the raw data for the real-space spin and chirality configurations obtained by the simulated annealing in Figs.~\ref{fig: spin1}, \ref{fig: spin2}, \ref{fig: spin3}, \ref{fig: spin4}, \ref{fig: spin5}, and \ref{fig: spin6}.

When $H=0$, the ground state becomes the $n_{\rm sk}=2$ F-SkX phase for small $J^{\rm inter}$, as shown in Fig.~\ref{fig: PD}(b). 
The real-space spin configurations in each sublattice are shown in the left panel of Fig.~\ref{fig: spin1}, where all the layers exhibit noncoplanar spin configurations. 
As shown in Figs.~\ref{fig: mq_0.05}(e) and \ref{fig: mq_0.05}(f), this state has equal intensities at $\bm{Q}_1$, $\bm{Q}_2$, and $\bm{Q}_3$ in magnetic moments; this state corresponds to a triple-$Q$ state.  
The difference between them is found in the positions of the vortex core, which are located around $S^z_i = +1$ in the left panel of Fig.~\ref{fig: spin1} so that the energy by the antiferromagnetic interlayer exchange interaction is gained. 
Reflecting the noncoplanar spin textures, there is a net scalar chirality, as shown in Figs.~\ref{fig: mq_0.05}(c) and \ref{fig: mq_0.05}(d), and the right panel of Fig.~\ref{fig: spin1}; the skyrmion number $n^{\eta}_{\rm sk}$ for layer $\eta$ is quantized as $+2$. 
This is also understood from the vorticity around the vortex core at $S^z_i = +1$ in a real-space picture (Fig.~\ref{fig: spin1}), where the in-plane spin components rotate two times around the core. 
Since all the layers exhibit the skyrmion number of two, we call this state $n_{\rm sk}=2$ F-SkX. 
It is noted that the state with $(n^{\rm A}_{\rm sk}, n^{\rm B}_{\rm sk}, n^{\rm C}_{\rm sk})=(-2,-2,-2)$ is degenerate with that with $(n^{\rm A}_{\rm sk}, n^{\rm B}_{\rm sk}, n^{\rm C}_{\rm sk})=(2,2,2)$. 

The appearance of the $n_{\rm sk}=2$ F-SkX is attributed to the positive biquadratic interaction $K$, which tends to favor multiple-$Q$ states compared to the single-$Q$ spiral state. 
Indeed, this state remains stable for infinitesimally small $J^{\rm inter}$; the model is connected to the single-layer triangular-lattice model for $J^{\rm inter} \to 0$, where the instability toward the $n_{\rm sk}=2$ SkX was reported in itinerant electron models, such as the Kondo lattice model~\cite{Ozawa_PhysRevLett.118.147205, Hayami_PhysRevB.99.094420, hayami2020multiple, hayami2021phase} and the Hubbard model~\cite{Kobayashi_PhysRevB.106.L140406}. 
Since the model Hamiltonian has rotational symmetry in spin space, the helicity and vorticity of the SkX are arbitrary taken; the SkX with $n_{\rm sk}=+2$ has the same energy as that with $n_{\rm sk}=-2$, whose degeneracy is lifted by considering an additional anisotropic exchange interaction originating from the spin--orbit coupling~\cite{Hayami_doi:10.7566/JPSJ.89.103702}. 
Similar SkXs with $n_{\rm sk}=+2$ have been also found in the model with anisotropic exchange interactions~\cite{amoroso2020spontaneous, yambe2021skyrmion, amoroso2021tuning, hayami2022multiple}. 

By increasing $J^{\rm inter}$, the $n_{\rm sk}=2$ F-SkX turns into the $n_{\rm sk}=2$ Ferri-SkX, as shown in the phase diagram in Fig.~\ref{fig: PD}(b).  
The real-space spin and chirality configurations of the $n_{\rm sk}=2$ Ferri-SkX are shown in the left and right panels of Fig.~\ref{fig: spin2}, respectively. 
In contrast to the $n_{\rm sk}=2$ F-SkX, the sign of the skyrmion number for one of three layers is reversed in order to further gain the energy by $J^{\rm inter}$; see the case with $n^{\rm C}_{\rm sk}=-2$ in Fig.~\ref{fig: spin2}.  
Thus, the alignment of the skyrmion number as $(n_{\rm sk}^{\rm A}, n_{\rm sk}^{\rm B}, n_{\rm sk}^{\rm C})=(+2, +2, -2)$ or $(-2,-2,+2)$ corresponds to the ferri-type one; the total skyrmion number is smaller than that in the $n_{\rm sk}=2$ F-SkX. 
Similarly, the total scalar chirality becomes small compared to that in the $n_{\rm sk}=2$ F-SkX, as shown in Figs.~\ref{fig: mq_0.6}(c) and \ref{fig: mq_0.6}(d).
Reflecting the different skyrmion numbers, three-sublattice spin configurations are not equivalent to each other; the intensities for the triple-$Q$ structures are anisotropic for two out of three sublattices, while those are almost isotropic for the remaining one, as shown in Figs.~\ref{fig: mq_0.6}(e) and \ref{fig: mq_0.6}(f).

Next, let us turn on the magnetic field $H$. 
We first discuss the case for small $J^{\rm inter}$ by taking $J^{\rm inter}=0.05$ in Fig.~\ref{fig: mq_0.05}; the zero-field state is the $n_{\rm sk}=2$ F-SkX. 
As $H$ increases, the magnetization $M^z_\eta$ gradually increases, as shown in Fig.~\ref{fig: mq_0.05}(b), while the scalar chirality remains almost the same, as shown in Fig.~\ref{fig: mq_0.05}(d). 
With a further increase of $H$, the $n_{\rm sk}=2$ F-SkX is replaced by the AF-SkX at $H\sim 0.18$ with jumps of the magnetization and scalar chirality for two out of three sublattices, as shown in Figs.~\ref{fig: mq_0.05}(b) and \ref{fig: mq_0.05}(d). 
The real-space spin and scalar chirality of the AF-SkX are shown in Fig.~\ref{fig: spin3}, where the skyrmion numbers for sublattices A, B, and C are given by $(n_{\rm sk}^{\rm A}, n_{\rm sk}^{\rm B}, n_{\rm sk}^{\rm C})=(1,-2,1)$. 
Note that the state with $(n_{\rm sk}^{\rm A}, n_{\rm sk}^{\rm B}, n_{\rm sk}^{\rm C})=(-1,2,-1)$ is degenerate with that with  $(n_{\rm sk}^{\rm A}, n_{\rm sk}^{\rm B}, n_{\rm sk}^{\rm C})=(1,-2,1)$; both of them are obtained by the simulated annealing at equal probability. 

By closely looking at the spin configuration in the left panel of Fig.~\ref{fig: spin3}, one finds that the skyrmion cores denoted by $S^z_i=-1$ for sublattice B are surrounded by the vortex with the winding number of $+2$. 
In contrast, the skyrmion cores for sublattices A and C are surrounded by the anti-vortex with the winding number of $-1$. 
Reflecting such a difference, the skyrmion number for sublattice B is -2 times that for sublattices A and C. 
Thus, the total skyrmion number in the magnetic unit cell becomes zero, which means the emergence of the AF-SkX. 
Nevertheless, it is noteworthy that the total scalar chirality $\chi^{\rm sc}_{\rm total}=\chi^{\rm sc}_{\rm A}+\chi^{\rm sc}_{\rm B}+\chi^{\rm sc}_{\rm C}$ becomes nonzero, as found from the data in Fig.~\ref{fig: mq_0.05}(d). 
This might be attributed to the nature of the discrete-lattice system. 
In the end, the contribution of the topological Hall effect should be finite even in the AF-SkX. 
Although the spin configuration for sublattice B is qualitatively different from that for sublattices A and C in the left panel of Fig.~\ref{fig: spin3}, both of them exhibit similar triple-$Q$ structures with almost the same intensity, as shown in Figs.~\ref{fig: mq_0.05}(e) and \ref{fig: mq_0.05}(f). 

By further increasing $H$, the AF-SkX changes into the 3$Q$ state, whose spin and chirality configurations in real space are shown in the left and right panels of Fig.~\ref{fig: spin4}, respectively.   
This state is mainly characterized by the double-$Q$ in-plane and single-$Q$ out-of-plane spin modulations with different intensities, as shown in Figs.~\ref{fig: mq_0.05}(e) and \ref{fig: mq_0.05}(f). 
Although the spin texture in the 3$Q$ state is noncoplanar, this state does not have a net scalar chirality in the whole system in contrast to the above SkXs in Figs.~\ref{fig: mq_0.05}(c) and \ref{fig: mq_0.05}(d); the scalar chirality oscillates along the direction where the out-of-plane spin component oscillates~\cite{Hayami_PhysRevB.94.174420}, as shown in the right panel of Fig.~\ref{fig: spin4}, and hence, the skyrmion number becomes zero.  

The 3$Q$ state shows the phase transition to the $n_{\rm sk}=1$ F-SkX as $H$ increases, as shown in Fig.~\ref{fig: PD}(b). 
This state consists of the SkX with the skyrmion number of one in each layer, as shown in the left panel of Fig.~\ref{fig: spin5}. 
The skyrmion cores are located at the interstitial site rather than the lattice site~\cite{Hayami_PhysRevResearch.3.043158}, and their positions for different sublattices are different from each other; the ABC stacking is realized in order to gain the energy by $J^{\rm inter}$~\cite{Lin_PhysRevLett.120.077202}. 
This state is energetically degenerate to the state with $(n_{\rm sk}^{\rm A}, n_{\rm sk}^{\rm B}, n_{\rm sk}^{\rm C})=(-1,-1,-1)$. 
When $H$ is increased into the $n_{\rm sk}=1$ F-SkX phase, the state turns into the 3$Q$ state again, and it continuously turns into the fully-polarized state, as shown in Figs.~\ref{fig: mq_0.05}(a) and \ref{fig: mq_0.05}(c). 

For large $J^{\rm inter}$, the $n_{\rm sk}=2$ F-SkX and the 3$Q$ state in the low-field region is no longer stabilized; the $n_{\rm sk}=2$ F-SkX is replaced by the $n_{\rm sk}=2$ Ferri-SkX, as discussed above. 
In addition, the Ferri-SkX II appears in the phase diagram between the $n_{\rm sk}=2$ Ferri-SkX and the AF-SkX, as shown in Fig.~\ref{fig: mq_0.6}.
As shown by the real-space spin configuration in the left panel of Fig.~\ref{fig: spin6}, the Ferri-SkX II exhibits distinct spin configurations for all the sublattices; the spin configuration for sublattice A consists of the SkX with the skyrmion number of $+2$, that for sublattice B consists of the SkX with the skyrmion number of $+1$, and that for sublattice C consists of the SkX with the skyrmion number of $-2$, where the real-space scalar chirality distributions are shown in the right panel of Fig.~\ref{fig: spin6}.  
Thus, the skyrmion number in this state is characterized by $+1$ in the whole layer, which means the emergence of a ferri-type alignment of the SkX, although the sequence of the skyrmion number in the three sublattices is different from that of the $n_{\rm sk}=2$ Ferri-SkX.
The other phase sequence in the high-field region is similar to that for small $J^{\rm inter}$ except for $0.85\lesssim J^{\rm inter} \lesssim 1$, where another phase denoted as $3Q$ II appears in the narrow region between the AF-SkX and $n_{\rm sk}=1$ F-SkX. 
This phase is characterized by an anisotropic triple-$Q$ structure without the skyrmion number for sublattices A--C (not shown). 

The appearance of various SkXs in the phase diagram in Fig.~\ref{fig: PD} is attributed to the sublattice degree of freedom in the lattice structure. 
For $J^{\rm inter}=0$ corresponding to the single-sublattice system~\cite{hayami2020multiple}, there are two types of SkXs in the phase diagram: $n_{\rm sk}=2$ F-SkX in the low-field region and $n_{\rm sk}=1$ F-SkX in the intermediate-field region. 
Meanwhile, additional three types of SkXs are realized by taking into account the interaction between different sublattices: $n_{\rm sk}=2$ Ferri-SkX, Ferri-SkX II, and AF-SkX. 
In these SkXs, the skyrmion number in each sublattice is different from each other, which indicates that the sublattice degree of freedom becomes a source of further rich topological spin textures as also found in the multi-sublattice SkXs in the two-sublattice honeycomb structure~\cite{Yambe_PhysRevB.107.014417}.

\section{Summary}
\label{sec: Summary}

To summarize, we have investigated the stability of the SkXs with an emphasis on the sublattice degree of freedom in the lattice structure. 
The ground-state phase diagram was constructed for the effective spin model on the three-sublattice triangular lattice by performing the simulated annealing. 
We have shown that two types of Ferri-SkXs and one type of AF-SkX are stabilized by the interplay between the biquadratic interaction and antiferromagnetic inter-sublattice interaction. 
The present results indicate that further composite SkXs consisting of sublattice-dependent skyrmion numbers are expected. 
One of the potential situations is a multi-layer system, where the SkX is stabilized in each layer, as demonstrated in this study. 
Once the interlayer interaction is antiferromagnetic, rich Ferri- and AF-SkXs can be realized.

\begin{acknowledgments}
This research was supported by JSPS KAKENHI Grants Numbers JP21H01037, JP22H04468, JP22H00101, JP22H01183, JP23K03288, JP23H04869, and by JST PRESTO (JPMJPR20L8). 
Parts of the numerical calculations were performed in the supercomputing systems in ISSP, the University of Tokyo.
\end{acknowledgments}


\bibliographystyle{apsrev}
\bibliography{../ref}
\end{document}